\documentclass[preprint,twocolumn,10pt,aps,longbibliography]{revtex4-2}
\usepackage[]{natbib}

\usepackage{graphicx}
\usepackage{dcolumn}
\usepackage{bm}
\usepackage[T2A]{fontenc}
\usepackage{comment}
\usepackage{braket}
\usepackage{amsmath}
\usepackage{amssymb}
\usepackage{dsfont}
\usepackage{upgreek}
\usepackage{natbib}
\newcommand{\bfp}{{\bf p}}
\newcommand{\bfx}{{\bf x}}
\newcommand{\bfq}{{\bf q}}
\newcommand{\bfk}{{\bf k}}
\newcommand{\bfqq}{{\bf q'}}
\DeclareMathOperator{\Tr}{Tr}
\DeclareMathOperator{\diag}{diag}
\DeclareMathOperator{\const}{const}

\begin{document}

\title{Generalized Lindblad master equation for neutrino evolution}

\author{Konstantin Stankevich}
\email{kl.stankevich@physics.msu.ru}
\affiliation{%
	Faculty of Physics,
	Lomonosov Moscow State University, Moscow 119992, Russia
}
\affiliation{%
	Institute for Nuclear Research of the Russian Academy of Sciences,
    Moscow 117312, Russia
}


\author{Alexander Studenikin}
\email{studenik@srd.sinp.msu.ru}
\affiliation{%
	Faculty of Physics,
	Lomonosov Moscow State University, Moscow 119992, Russia
}
\affiliation{%
	Institute for Nuclear Research of the Russian Academy of Sciences,
    Moscow 117312, Russia
}

\author{Maksim Vyalkov}
\email{vyalkovmsu@yandex.ru}
\affiliation{%
	Faculty of Physics,
	Lomonosov Moscow State University, Moscow 119992, Russia
}
\affiliation{%
	Branch of Lomonosov Moscow State University in Sarov, 607328 Russia
}
\affiliation{%
	National Centre for Physics and Mathematics, Sarov 607182, Russia
}

\date{\today}

\begin{abstract}

A new theoretical framework, based on the quantum field theory of open systems applied to neutrinos, has been developed. This framework aims to describe the neutrino evolution in external environment, taking into account the effect of neutrino quantum decoherence. We have applied this approach to investigate a novel mechanism for neutrino quantum decoherence, which arises due to the neutrino decay into a lighter neutrino state and a massless particle, as well as the inverse process of absorption of a massless particle by a neutrino. 
We derived the new generalized Lindblad master equation for the neutrino evolution that accounts for neutrino transitions between states with different momenta. We also demonstrate that studying of neutrino quantum decoherence through this master equation provides a unique possibility to determine  or limit the neutrino decay width. On this basis we  obtained the constraint on neutrino lifetime $\frac{\tau_2}{m_2} > 1.83 \times 10^{-10} \frac{s}{eV}$ (for the case of the Dirac neutrino scalar decay and neutrino degenerate hierarchy).

\end{abstract}

\pacs{Valid PACS appear here}
\maketitle


\section{\label{sec:level1} Introduction}

Neutrinos have provided a window to new physics for nearly a century \cite{Giunti:2014ixa}. One of the promising areas of research in neutrino physics is the search for evidence of the effect of neutrino quantum decoherence, which can occur as a result of neutrino interaction with an external matter. This phenomenon can be described by the Lindblad equation \cite{Lindblad:1975ef,Gorini:1975nb}

\begin{multline} \label{Lindblad_eq}
    \dfrac{\partial \rho_\nu(t)}{\partial t} = - i \left[  H(t),\rho_\nu(t) \right] 
    + \\ + 
    \sum_i \gamma_i\left(L_i \rho_\nu L_i^{\dagger}-\frac{1}{2}\left\{L_i^{\dagger} L_i, \rho_\nu\right\}\right)
    ,
\end{multline}
where $\rho_\nu(t)$  is the neutrino density matrix, $H(t)$ is the Hamiltonian, $L_i$ and $\gamma_i$ are Lindblad dissipative operators and parameters, that describe the effect of neutrino quantum decoherence. Note that the Lindblad equation defines only the structure of the neutrino evolution equation, but not the form of dissipative operators or values of parameters. It is also worth noting that the Lindblad equation can be used to describe quantum decoherence that occurs due to transitions between particle's different states with the same momentum, which is not always the case for a moving neutrino. The aim of the neutrino experimental data analysis is to find constraints on the decoherence parameters associated with various configurations of the dissipative operators. The significance of the investigation of neutrino quantum decoherence stems from the fact that this phenomenon can be caused by physics beyond the Standard model, and its detection could provide indications of new physics. 

The search for evidence of the effect of neutrino quantum decoherence is conducted in neutrino fluxes from various sources. In this context, we would like to highlight several papers on searching for the effect of quantum decoherence in neutrino fluxes from nuclear reactors and accelerators \cite{BalieiroGomes:2016ykp, Oliveira:2016asf, Coelho:2017byq, deGouvea:2021uvg, deGouvea:2020hfl, DeRomeri:2023dht}, the sun \cite{deHolanda:2019tuf,Fogli:2007tx} and in atmospheric neutrino fluxes \cite{Coloma:2018idr}. Recently it was shown that the neutrino quantum decoherence may provide an explanation for the gallium neutrino anomaly \cite{Farzan:2023fqa}. 

Recent studies \cite{Purtova:2023hcy, Stankevich:2023qdi, Stankevich:2020sja, dosSantos:2023skk} have demonstrated that neutrino quantum decoherence may influence neutrino fluxes from supernova bursts, indicating that it is feasible to search for a corresponding effect in supernovae neutrino fluxes as well. In particular, the impact of quantum decoherence on collective neutrino oscillations has been explored in \cite{Purtova:2023hcy, Stankevich:2023qdi, Stankevich:2020sja}.

Nevertheless, the mechanisms underlying the occurrence of neutrino quantum decoherence are currently not well studied. In \cite{Burgess:1996mz,PhysRevD.71.013003, PhysRevD.50.4762} the neutrino quantum decoherence due to classical fluctuations of the external environment was considered. Here, we would like to draw attention to  \cite{PhysRevD.104.043018}, where it is demonstrated that neutrino quantum decoherence can occur as a result of spacetime fluctuations. In \cite{Nieves:2020jjg,Nieves:2019izk}, a study on the mechanism of quantum decoherence resulting from the neutrino scattering has been presented.

In our study \cite{PhysRevD.101.056004} we propose a novel approach to the description of neutrino quantum decoherence based on the quantum electrodynamics of open systems \cite{Breuer:2007juk,brasil2013simple}. Within this approach it is shown that the neutrino radiative decay under extreme astrophysical conditions results in neutrino quantum decoherence.

In this paper, we develop our approach to describe neutrino evolution accounting for neutrino arbitrary decay accompanied by the emission of a massless particle and the reverse process of absorption of a massless particle by neutrino as well.

It is worth noting that the investigation of the neutrino decay is a significant area of research in neutrino physics. In particular, active research is underway to investigate the radiation decay of neutrino \cite{Ternov:2014wra,Ternov:2013ana,Giunti:2014ixa}, including the studies of a novel radiation phenomenon known as neutrino spin light \cite{Grigoriev:2017wff}. Neutrino physics also focuses on the neutrino decay into scalar and pseudo-scalar particles \cite{Funcke:2019grs}. Additionally, neutrino may decay into gravitons while propagating through a medium \cite{Nieves:2009by}. 


\section{Evolution master equation}

We consider a Dirac neutrino, which can be described by a density matrix

\begin{equation}
    \rho_\nu (t) = \ket{\Phi(t)} \bra{\Phi(t)}
    ,
\end{equation}
where the state $\ket{\Phi(t)}$ describes neutrino with different momenta, i.e.

\begin{equation}
    \ket{\Phi}=\underset{\mathbf{p}}{\otimes} \ket{\Phi_{\mathbf{p}}}
    .
\end{equation}
The aim of our study is to derive the neutrino evolution equation that accounts for transitions between different stationary neutrino states, accompanied by the emission or absorption of a massless particle (i.e. taking into account the decay of a neutrino into a lighter neutrino state and a massless particle and the reverse absorption process of a massless particle by a neutrino).  The stationary neutrino state $i$ with momentum $\bfp$ is given by 

\begin{equation}
\ket{i \bfp} = \ket{i} \otimes \ket{\bfp} = \sqrt{2 E_{\bfp i}} a_{\bfp i}^{\dagger} \ket{0}.
\end{equation}
Here $a_{\bfp i}$ and $a_{\bfp i}^{\dagger}$ are the annihilation and creation operators of neutrino. The stationary state $i$ also takes into account the helicity of the particle. The matrix elements of the density matrix are written in the form

\begin{equation}
\rho_\bfp^{ij}(t) =  \bra{i \bfp} \rho_\nu (t) \ket{j \bfp} 
.
\end{equation}
These matrix elements coincide with the elements of the density matrix defined in \cite{Sigl:1993ctk} (see also \cite{Kharlanov:2020cti}), if one considers the stationary states to be neutrino mass states. It is assumed  that there is no superposition between states with different momentum in the system, i.e. the non-diagonal momentum elements of the density matrix are zero

\begin{equation}\label{diag_p}
    \bra{i \bfp} \rho_\nu(t) \ket{j \bfq} =  (2 \pi)^3 \delta^{(4)}(\bfp-\bfq) \rho_\bfp^{ij'}(t)
    .
\end{equation}
Transitions between neutrino and antineutrino are not taken into account in this study. Therefore, all terms related to antineutrino will be omitted in the following discussion.

It is assumed that the neutrinos are in a bath of massless particles in equilibrium. Then the evolution of the density matrix for the entire system $\varrho(t)$ (neutrinos + massless particles) can be written in the interaction representation using the evolution operator 
$U(t_0, t)$ 

\begin{equation}\label{}
	\varrho \left( t\right) = U(t_0, t) \: \varrho (t_0) \: U^{\dagger}(t_0, t).
\end{equation}
It is also assumed that at the initial point in time, the systems are weakly coupled, i.e., the density matrix for the entire system can be expressed as a direct product of the density matrices of the neutrinos $\rho_\nu$ and the external environment $\rho_A$

\begin{equation}
    \varrho(t_0) = \rho_\nu (t_0) \otimes \rho_A (t_0)
    .
\end{equation}
The evolution operator is given in the following form

\begin{equation}
    U(t_0, t) = Texp \left( -i \int^t_{t_0} H(t') dt' \right)
    .
\end{equation}
The Hamiltonian $H(t)$ that describes the interaction between neutrino and an external medium composed of massless particles can be written in the most general form

\begin{equation} \label{Hamiltonian}
	H(t) = \int d^3 \bfx \sum_{\alpha} j_{\alpha}(x) A_{\alpha}(x)
	,
\end{equation}
where $A_\alpha$ is a massless field, $j_\alpha = \bar {\nu}(x) \Gamma_\alpha \nu (x)$ is the neutrino current, $\Gamma_\alpha$ is the interaction vertex that includes the interaction constant. In the case of neutrino radiation decay, the vertex is replaced by a general electromagnetic vertex $\Gamma_\alpha \to \Gamma_{\mu}$, and the field is replaced by electromagnetic field $A_\alpha(x) \to A_{\mu}(x)$ \cite{Giunti:2014ixa}. In the case of the decay of a neutrino into a massless scalar particle (such as an axion), it is necessary to consider the replacement $A_\alpha(x) \to \phi (x) $ for the field and $\Gamma_\alpha \to (g^s\mathds{1} + i g^p \gamma^5)$ for a scalar and pseudoscalar vertex \cite{Funcke:2019grs,Moss:2017pur}. 
The interaction of neutrinos with gravitons or massless particles of higher spins can also be considered: $H = j_{\mu\hdots\nu} A^{\mu\hdots\nu}$. 

Next, we consider a general form of the Hamiltonian, which corresponds to one of the three possible interactions. The neutrino wave function operator can be expressed as

\begin{equation}
    \nu (x) =  \sum_i \int \dfrac{d^3 \bfp }{(2 \pi)^3} \dfrac{1}{\sqrt{2 E_{\bfp i}}} a_{{\bf p} i} u_i(\bfp) e^{-i p x}
    .
\end{equation}
Here, the negative-frequency solutions with the antineutrino creation operator $b_{{{\bf p} i}}^\dagger$ were omitted. 

The expansion of the evolution operator $U(t)$ in terms of powers of the coupling constant is (only the first two terms are of interest for our study)

\begin{widetext}

\begin{equation}\label{key}
	U\left(t, t_0\right)=1+(-i) \int_{t_0}^t d t_1 H\left(t_1\right)+(-i)^2 \int_{t_0}^t d t_1 \int_{t_0}^{t_1} d t_2 H\left(t_1\right) H\left(t_2\right) + ...
\end{equation}
Using this expansion, the density matrix takes the form

\begin{equation}\label{varrho}
    \varrho \left( t\right) =  \varrho(t_0) - i \int_{t_0}^t\left[H, \varrho(t_0)\right] d t   
    + 2 \int_{t_0}^t d t_1 \int_{t_0}^{t_1} d t_2 \: \: H(t_1) \varrho(t_0) H(t_2) 
    - \int_{t_0}^{t}  d t_1 \int_{t_0}^{t1} dt_2 \: \: \{ H_I\left(t_1\right) H_I\left(t_2\right), \varrho(t_0)  \}
    .
\end{equation}
\end{widetext}
The evolution of the environment is not under our consideration, therefore it is necessary to trace out its degrees of freedom  (in this case, the left-hand side of the equation gives a density matrix that describes neutrinos $\rho_\nu (t) =  \Tr_A \varrho (t)$). Taking the trace and the time derivative of the equation (\ref{varrho}) and using the explicit form of the Hamiltonian (\ref{Hamiltonian}) we obtain

\begin{equation}\label{equation_for_rho}
\begin{aligned}
    \dfrac{\partial \rho_\nu (t)}{\partial t} =  - i \Tr_a  \int d^3 \bfx \left[j_\alpha(x) \braket{A^\alpha(x)}, \rho_\nu(t_0) \right] + \\
    + 2 \int d^3 \bfx_1 \int_{t_0}^{t_1} d^4 x_2 D_+^{\alpha \beta} (x_1-x_2) j_\alpha(x_2) \rho_\nu(t) j_\beta(x_1) - \\
    - \int d^3 \bfx_1 \int_{t_0}^{t_1} d^4 x_2 D_+^{\alpha \beta} (x_1-x_2) \{ j_\alpha(x_1) j_\beta(x_2), \rho_\nu(t) \}
    ,
\end{aligned}
\end{equation}
where the angle brackets denote the averaging and $\braket{A_\alpha} = \Tr_A A = 0$. If there is a classical field present in the system, this term will not be equal to zero. For example, if there is a classical magnetic field and a field of photons, then the field $A^\alpha$ can be decomposed into a classical and quantum component $A^\alpha = A^\alpha_{cl} + A^\alpha_{q}$. In this case, the average is $\braket{A^\alpha} = \braket{A^\alpha_{cl}} \neq 0$ and gives the classical magnetic component. In further calculations, this term is omitted.

The  definition for the correlation function is introduced 

\begin{equation}
D_+^{\alpha \beta}(x)=\Tr_A\left[A^\alpha(x_1) A^\beta(x_2)\right] = \braket{A^\alpha(x_1) A^\beta(x_2)}.    
\end{equation}
The field of a massless particles can be represented as

\begin{equation}\label{key}
    A_{a}(x)=\int \frac{d^3 k}{\sqrt{2(2 \pi)^3 \omega}} \sum_{\lambda=1,2}\left[e_{a}^{\lambda}(k) b^{\lambda}_k e^{-i k x}+e_a^{\lambda}(k) b^{\lambda \dagger}_k e^{+i k x} \right]
    ,
\end{equation}
where $b^{\lambda \dagger}_k$ and $b^{\lambda}_k$ are the creation and annihilation operators and $e_{a}^{\lambda}(k)$ is the polarization of the massless particle. The averaging should be carried out according to the equilibrium state of the external environment 

\begin{equation}
    \langle O \rangle = tr_a\left(O \frac{1}{Z} \exp [-H_a/T]\right)
    ,
\end{equation}
where $H_a$ is the free Hamiltonian of the particles of the medium and $T$ is the temperature. The rules of averaging the creation and annihilation operators are

\begin{equation}\label{corell}
	\begin{aligned}
		\left\langle b_\lambda(k) b_{\lambda^{\prime}}\left(k^{\prime}\right)\right\rangle & =\left\langle b_\lambda^{\dagger}(k) b_{\lambda^{\prime}}^{\dagger}\left(k^{\prime}\right)\right\rangle=0, \\
		\left\langle b_\lambda(k) b_{\lambda^{\prime}}^{\dagger}\left(k^{\prime}\right)\right\rangle & =\delta_{k k^{\prime}} \delta_{\lambda \lambda^{\prime}}\left(1+N\left(\omega\right)\right), \\
		\left\langle b_\lambda^{\dagger}(k) b_{\lambda^{\prime}}\left(k^{\prime}\right)\right\rangle & =\delta_{k k^{\prime}} \delta_{\lambda \lambda^{\prime}} N\left(\omega\right),
	\end{aligned}
\end{equation} 
where

\begin{equation}\label{key}
    N\left(\omega\right)  =\frac{1}{\exp \left[ \omega/T \right]-1}
\end{equation}
is the distribution of massless particles. Finally, the correlation function is obtained in the form

\begin{equation}\label{key}
    \begin{aligned}
        &D_{+}\left(x,x^{\prime}\right)_{\alpha \beta} =	\left\langle A_{\alpha}(x)A_{\beta}(x^{ \prime})\right\rangle= \\
        &= \int \frac{d^3 \bfk}{2(2 \pi)^3 \omega} P_{\alpha \beta} \left(  \left(1+N\left(\omega\right)\right) e^{-ik\left(  x - x^{\prime}\right) } + N\left(\omega\right) e^{ik\left( x-  x^{\prime}   \right) }   \right) 
        ,
    \end{aligned}
\end{equation} 
where  $ P_{\alpha \beta} $ is given by
\begin{equation}\label{key}
    P_{\alpha \beta} \equiv  \sum_{\lambda } e_{\alpha}^{\lambda}(k)e_{\beta}^{\lambda}(k) 
    .
\end{equation}

To obtain a Lindblad-type equation, it is necessary, at first, to make an approximation of a rotating wave. To do this approximation it is convenient to write down equation (\ref{equation_for_rho}) in terms of the stationary neutrino states.

\begin{widetext}
\begin{equation}\label{equation_1}
\begin{aligned}
    \dfrac{\partial \rho^{ij}_\bfp(t)}{\partial t} = 
    2 \sum_{n,n'} \int d^3 q \int d^3 q'  \int d^3 \bfx_1 \int_{t_0}^{t_1} d^4 x_2 D_+^{\alpha \beta} \bra{i \bfp}j^{(2)}_\alpha \ket{n \bfq} \bra{n \bfq} \rho_\nu(t) \ket{n' \bfqq} \bra{n' \bfqq} j^{(1)}_\beta \ket{j \bfp} - 
    \\
    - \sum_{n,n'} \int d^3 q \int d^3 q' \int d^3 \bfx_1 \int_{t_0}^{t_1} d^4 x_2 D_+^{\alpha \beta}  \bra{i \bfp} j_\alpha^{(1)} \ket{n \bfq} \bra{n \bfq} j_\beta^{(2)} \ket{n' \bfqq} \bra{n' \bfqq} \rho_\nu(t) \ket{j \bfp} -
    \\
    - \sum_{n,n'} \int d^3 q \int d^3 q' \int d^3 \bfx_1 \int_{t_0}^{t_1} d^4 x_2 D_+^{\alpha \beta} \bra{i \bfp} \rho_\nu(t) \ket{n \bfq} \bra{n \bfq} j_\beta^{(2)} \ket{n' \bfqq} \bra{n' \bfqq} j_\alpha^{(1)}  \ket{j \bfp}
    ,
\end{aligned}
\end{equation}
\end{widetext}
where it is defined $D_+^{\alpha \beta} = D_+^{\alpha \beta}(x_1-x_2)$, 
$j_\alpha^{(1)} = j_\alpha(x_1)$, 
$j_\alpha^{(2)} = j_\alpha(x_2)$, 
$\int d^3 q  = \int \dfrac{d^3 \bfq}{2 (2\pi)^3 E_{\bfq n}}$,
$\int d^3 q'  = \int \dfrac{d^3 \bfqq}{2 (2\pi)^3 E_{\bfqq n'}}$. The matrix elements of the neutrino current are expressed as

\begin{equation}
    \bra{i \bfp}j_\alpha(x) \ket{j \bfp} = \bar{u}_{i}(\bfp) \Gamma_\alpha u_j(\bfp) e^{i (p-q) x }
    .
\end{equation}
It can be seen that the matrix element is proportional to the exponent $e^{i(E_p - E_q) t}$. The rotating-wave approximation in the context of the quantum optical regime (also known as the secular approximation) involves retaining only the secular components in the double sums over the system frequencies and is equivalent to an averaging process for rapidly oscillating terms. In order to accomplish this, the Kronecker delta symbols should be added to equation (\ref{equation_1})

\begin{widetext}
\begin{equation}\label{equation_2}
\begin{aligned}
    \dfrac{\partial \rho^{ij}_\bfp(t)}{\partial t} 
    + 2 \sum_{n,n'} \int d^3 q \int d^3 q'  \int_{t_0}^t d^4 x_1 \int_{t_0}^{t_1} d^4 x_2 D_+^{\alpha \beta} \bra{i \bfp}j^{(2)}_\alpha \ket{n \bfq} \rho_\bfq^{nn'}(t) \bra{n' \bfqq} j^{(1)}_\beta \ket{j \bfp} \delta_{nn'} \delta_{ij} (2\pi)^3 \delta^{(4)}(\bfq-\bfqq) -
    \\
    - \sum_{n,n'} \int d^3 q \int d^3 q' \int_{t_0}^t d^4 x_1 \int_{t_0}^{t_1} d^4 x_2 D_+^{\alpha \beta}  \bra{i \bfp} j_\alpha^{(1)} \ket{n \bfq} \bra{n \bfq} j_\beta^{(2)} \ket{n' \bfqq} \rho_\bfp^{n'j}(t) \delta_{in'} (2\pi)^3 \delta^{(4)}(\bfp-\bfqq) -
    \\
    - \sum_{n,n'} \int d^3 q \int d^3 q' \int_{t_0}^t d^4 x_1 \int_{t_0}^{t_1} d^4 x_2 D_+^{\alpha \beta} \rho_\bfp^{in}(t) \bra{n \bfq} j_\beta^{(2)} \ket{n' \bfqq} \bra{n' \bfqq} j_\alpha^{(1)}  \ket{j \bfp} \delta_{jn} (2\pi)^3 \delta^{(4)}(\bfp-\bfq)
    ,
\end{aligned}
\end{equation} 
\end{widetext}
where it was taken into account the diagonality of the neutrino density matrix by momentum (\ref{diag_p}). In equation (\ref{equation_2}) there are integrals of the following form

\begin{equation}\label{int_time}
    \int d^3 \bfx_1 \int_{t_0}^{t_1} d^4 x_2 e^{i p x_1} e^{-i p x_2}
    .
\end{equation}
The delta function will arise after integrating along the three-dimensional coordinate axis. For the case of weak interaction between neutrinos and the external environment it is possible to make the Markovian approximation: the lower limit of the time integral can be replaced by $t_0 \rightarrow -\infty$. Then, after replacing $\tau = t_1 - t_2$, the integral (\ref{int_time}) is written as

\begin{multline}\label{markov}
   (2\pi)^3 \delta^{(3)}(\bfp) \int_0^\infty d \tau e^{-i E_p \tau} 
   = \\ = 
   \dfrac 1 2 (2 \pi)^4 \delta^{(4)}(p) - i (2 \pi)^3 \delta^{(3)}(\bfp) P \dfrac{1}{E_p}
   ,
\end{multline}
where in the second term $P$ means the Cauchy principal value (see for more details in \cite{brasil2013simple}). The second term has an imaginary value and therefore will contribute to the  Hamiltonian (i.e. to the coherent part of evolution), which is not the focus of this paper. Therefore, the part containing the imaginary unit will be omitted. 

Using the expressions (\ref{equation_2}) and (\ref{markov}) the final equation of neutrino evolution taking into account transitions between stationary states is obtained

\begin{widetext}
\begin{equation}\label{Main_Evoltion}
    \begin{aligned}
        \dfrac{\partial \rho_\bfp(t)}{\partial t} 
        =  
        \left[ H (t), \rho_\bfp(t) \right]
        - \dfrac 1 2 \sum_i  \left( \Gamma^d_{i\bfp} + \Gamma^a_{i\bfp}  \right)  \left\{ \Pi_{ii}, \rho_\bfp(t) \right\} +
        \\
        + \sum_i \int \dfrac{d^3 \bfk}{2 (2\pi)^3 \omega} \left[\sum_{j:\{m_j>m_i\}} \int \dfrac{d^3 \bfq}{2 (2\pi)^3 E_{\bfq j}} \Gamma^d_{j \bfq \to i \bfp}\Pi_{ij} \rho_\bfq(t) \Pi_{ji} 
        + 
        \sum_{j:\{m_j<m_i\}} \int \dfrac{d^3 \bfq}{2 (2\pi)^3 E_{\bfq j}} \Gamma^a_{j \bfq \to i \bfp}\Pi_{ij} \rho_\bfq(t) \Pi_{ji}\right]
        ,
    \end{aligned}
\end{equation}
where $\Pi_{ij} = \ket{i}\bra{j}$ is the projector on the neutrino stationary state, $\Gamma^d_{i \bfp}$ is the width of the decay of the neutrino stationary state $\ket{i \bfp}$ to all possible neutrino states and  $\Gamma^a_{i \bfp}$ is the width of the inverse process of transition from all possible states into the neutrino state $\Gamma^a_{i \bfp}$ (the inverse process is accompanied by absorption of a massless particle by the neutrino)    

\begin{equation}
    \Gamma^d_{i \bfp} = \sum_{j:\{m_j<m_i\}} \int \dfrac{d^3 \bfk}{2 (2\pi)^3 \omega} \int \dfrac{d^3 \bfq}{2 (2\pi)^3 E_{\bfq j}} \left[ 1+N(\omega) \right] \dfrac{(2 \pi)^4 \delta^{(4)}(p-q-k) P^{\alpha \beta} \left( \bar u_{\bfp i} \Gamma_\alpha u_{\bfq j} \bar u_{\bfq j} \Gamma_\beta u_{\bfp i} \right) } {2 E_{\bfp i}}
    ,
\end{equation}

\begin{equation}
\Gamma^a_{i \bfp} = \sum_{j:\{m_j>m_i\}} \int \dfrac{d^3 \bfk}{2 (2\pi)^3 \omega} \int \dfrac{d^3 \bfq}{2 (2\pi)^3 E_{\bfq j}} N(\omega) \dfrac{(2 \pi)^4 \delta^{(4)}(p+k-q) P^{\alpha \beta} \left(\bar u_{\bfq j} \Gamma_\alpha u_{\bfp i} \bar u_{\bfp i} \Gamma_\beta u_{\bfq j} \right) } {2 E_{\bfp i}}
.
\end{equation}
The widths of transition between neutrino states $\ket{j\bfq} \to \ket{i \bfp}$ accompanied by the massless particle emission $\Gamma^d_{j\bfq \to i \bfp}$ and massless particle absorption $\Gamma^a_{j\bfq \to i \bfp}$ are expressed as

\begin{equation}
\Gamma^d_{j\bfq \to i \bfp} = \left[ 1+N(\omega) \right] \dfrac{(2 \pi)^4 \delta^{(4)}(q-p-k) P^{\alpha \beta} \left(  \bar u_{\bfq j} \Gamma_\alpha u_{\bfp i} \bar u_{\bfp i} \Gamma_\beta u_{\bfq j} \right) } {2 E_{\bfq j}}
,
\end{equation}

\begin{equation}
\Gamma^a_{j\bfq \to i \bfp} = N(\omega) \dfrac{(2 \pi)^4 \delta^{(4)}(q+k-p) P^{\alpha \beta} \left(  \bar u_{\bfq j} \Gamma_\alpha u_{\bfp i} \bar u_{\bfp i} \Gamma_\beta u_{\bfq j} \right) } {2 E_{\bfq j}}
.
\end{equation}
\end{widetext}

The main result of our research is the neutrino evolution equation (\ref{Main_Evoltion}) that accounts for transitions between different neutrino stationary states with different momentum due to neutrino decay on a massless particle and due to inverse process of massless particle absorption by the neutrino. 

The derived equation takes the form of the Lindblad equation (\ref{Lindblad_eq}). However, previous studies on neutrino quantum decoherence that have used the Lindblad equation have not considered transitions between neutrinos with different momenta. 

Moreover, we have obtained the exact form of the dissipative operators and the exact meaning of the dissipative parameters. It gives opportunity to connect the obtained in experiments constraints on neutrino quantum decoherence with physical processes.

Additionally, it is worthwhile to compare our results with those obtained in \cite{Moss:2017pur}, where the evolution equation was derived by using a non-Hermitian Hamiltonian. First, the derived master equation (\ref{Main_Evoltion}) takes into account the process of massless particle absorption. Second, the third term on the right-hand side of equation (\ref{Main_Evoltion}) is absent comparing with \cite{Moss:2017pur} while the first two terms are identical.

\section{Neutrino decay on a scalar particle}

This section is dedicated to a special case of the obtained equation (\ref{Main_Evoltion}). This case demonstrates the possibility to find or constrain the neutrino lifetime in experimentrs wigth neutrino fluxes from accelerators and reactors. Namely, the effect of neutrino quantum decoherence, which arises from the decay of a neutrino mass state $\nu_j$ into a massless scalar particle and lighter neutrino mass state $\nu_i$ through a scalar interaction (for simplicity we use $m_3>m_2>m_1$), is considered. 

The neutrino interaction with a scalar particle is given by the Lagrangian 

\begin{equation}
    \mathcal{L}_{\text {int }}=i \phi \sum_{i \neq j} g_{i j} \bar{\nu}_i \nu_j+\text { H.c. }
    ,
\end{equation}
where $\phi$ is the scalar field, $g_{i j}$ is the interaction constant. In this section only the degenerate limit ($m_i\approx m_j$) is considered for which the flip of the neutrino spin  is suppressed \cite{Funcke:2019grs}. Therefore, it is enough to consider only three left-handed states  $\nu_1$, $\nu_2$ and $\nu_3$.

Neutrinos are an ultra relativistic particles. It means, firstly, that $E_\bfp \approx |\bfp|$ and, secondly, the scattering angle of neutrinos during decay is approximately zero. Therefore,  $\rho_\bfp(t) \equiv \rho(|\bfp|, \theta_\bfp,\phi_\bfp, t) \to \rho(|\bfp|,t)$ (where $\theta_\bfp$ and $\phi_\bfp$ are neutrino propagation angles). Finally, after integrating over angles in equation (\ref{Main_Evoltion})

\begin{widetext}
\begin{equation}\label{special_equation}
    \begin{aligned}
        \dfrac{\partial \rho (|\bfp|,t)}{\partial t} = - i \left[  H(t),\rho (|\bfp|,t) \right] + \\
        + \sum_{i,f:\{i>f\}} \left[ - \dfrac 1 2  \Gamma_{if}^{S} \left\{ \Pi_{ii}, \rho (|\bfp|,t) \right\} + g_{if}^2 \int_{|\bfp|}^{\left(\frac{m_i}{m_f}\right)^2 |\bfp|} \dfrac{d |\bfq|}{16 \pi^4} \left(  \dfrac{m_f^2}{|\bfp|^2} + \dfrac{m_i^2}{|\bfq|^2} + 2 \dfrac{m_i m_f}{|\bfp| |\bfq|} \right) \Pi_{if} \rho (|\bfq|,t) \Pi_{fi} \right]
        .
    \end{aligned}
\end{equation}
For the two-flavour approximation it is
\begin{equation}\label{special_equation_2}
    \dfrac{\partial \rho (|\bfp|,t)}{\partial t} = - i \left[  H(t),\rho (|\bfp|,t) \right] - \dfrac 1 2   \Gamma_{21}^{S} \left\{ \Pi_{22}, \rho (|\bfp|,t) \right\} + g_{12}^2 \int_{|\bfp|}^{\left(\frac{m2}{m1}\right)^2 |\bfp|} \dfrac{d |\bfq|}{16 \pi^4} \left(  \dfrac{m_1^2}{|\bfp|^2} + \dfrac{m_2^2}{|\bfq|^2} + 2 \dfrac{m_1 m_2}{|\bfp| |\bfq|} \right) \Pi_{21} \rho (|\bfq|,t) \Pi_{12}
    ,
\end{equation}

\end{widetext}
where it is supposed that neutrinos with different masses have the same momenta but different energies.  The integration in (\ref{special_equation}) takes place over the allowed energy range for the daughter neutrino \cite{Lindner:2001fx}. The decay width $\Gamma_{if}^{S}$ of the neutrino state $i$ to the neutrino state $f$ and a scalar particle in degenerate limit $m_i \approx m_f$ \cite{Funcke:2019grs} is
\begin{equation}
    \Gamma_{if}^{S} = \dfrac{g_{if}^2}{\pi} \dfrac{\left( m_i^2-m_f^2 \right)}{|\bfp|}
    .
\end{equation}
It is important to note that the number of neutrinos of a certain energy changes due the decay process, i.e. the trace of the neutrino density matrix $\Tr(\rho(|\bfp|,t)$ changes with time. However, the total number of neutrinos of all momenta remains the same $\int d |\bfp| \Tr(\rho(|\bfp|,t) = \const$.

In the degenerate case the integration goes over a small energy range $\left( \dfrac{m_i}{m_f} \approx 1\right)$. Therefore, it is possible to consider $\rho (|\bfq|,t)\approx \const$ and take the neutrino density matrix $\rho (|\bfq|,t)$ outside the integral sign. Then we get 

\begin{equation}
    g_{ij}^2 \int_{|\bfp|}^{\left(\frac{m_i}{m_f}\right)^2 |\bfp|} \dfrac{d |\bfq|}{16 \pi^4} \left(  \dfrac{m_f^2}{|\bfp|^2} + \dfrac{m_i^2}{|\bfq|^2} + 2 \dfrac{m_i m_f}{|\bfp| |\bfq|} \right) \approx \Gamma_{if}^{S}
    .
\end{equation}
Using this equality the master equation (\ref{special_equation}) is reduced to the usual Lindblad equation (\ref{Lindblad_eq}) with three dissipative operators $L_i$ and corresponding parameters $\gamma_i$:

\begin{equation}\label{dis_oper_param}
\begin{aligned}
    L_1=\Pi_{12} \ \ \ \ \text{\&} \ \ \ \ \gamma_1 = \Gamma_{21}^{S} ,  \\
    L_2=\Pi_{13} \ \ \ \ \text{\&}  \ \ \ \ \gamma_2 = \Gamma_{31}^{S} ,  \\
    L_3=\Pi_{23} \ \ \ \ \text{\&} \ \ \ \ \gamma_3 = \Gamma_{32}^{S} .
    \end{aligned}
\end{equation}
It is convenient to expand the Lindblad equation on Pauli matrices in the case of two neutrino oscillations or on Gell-Mann matrices in the case of three-flavour oscillations ($\rho(t) = \sum_n P_n F_n$ and $O= \sum_n O_n F_n$ where $F_n$ stands for Pauli or Gell-Mann matrices and $F_0$ is the unit matrix). Then the Lindblad equation can be written in the following form

\begin{equation} \label{3x3.eq.LindbladGM1}
\frac{\partial P_k (t)}{\partial t} F_k = 2\epsilon_{ijk} H_i P_j(t) F_k + D_{kl} P_k(t) F_l
,
\end{equation}
where the dissipative effects are encoded in the matrix $D_{kl}$ ( the indexes $k$ and $l$ starts with zero, i.e. $k,l = 0,1,2...$). For two-flavour oscillations (\ref{dis_oper_param}) the obtained dissipative matrix is

\begin{equation}
    D^{(2)}_{kl} = - \dfrac{\Gamma_{21}^{S}}{2}
\begin{pmatrix}
0 & 0 & 0 & 1  \\
0 & 1 & 0 & 0 \\
0 & 0 & 1 & 0 \\
0 & 0 & 0 & 2 
\end{pmatrix}
,
\end{equation}
and for the three-flavour case it is

\begin{widetext}
\begin{equation}\label{dis_matrix_3}
    D^{(3)}_{kl} = - \dfrac 1 2 
\begin{pmatrix}
0 & 0 & 0 & - \dfrac 1 3 \left( 2 \Gamma_{21}^{S} + \Gamma_{31}^{S} - \Gamma_{32}^{S} \right) & 0 & 0 & 0 & 0 & - \dfrac{1}{\sqrt 3} \left( \Gamma_{32}^{S} + \Gamma_{31}^{S} \right) \\
0 & \Gamma_{21}^{S} & 0 & 0 & 0 & 0 & 0 & 0 & 0 \\
0 & 0 & \Gamma_{21}^{S} & 0 & 0 & 0 & 0 & 0 & 0 \\
0 & 0 & 0 & 2 \Gamma_{21}^{S} & 0 & 0 & 0 & 0 & 0 \\
0 & 0 & 0 & 0 & \Gamma_{32}^{S} + \Gamma_{31}^{S} & 0 & 0 & 0 & 0 \\
0 & 0 & 0 & 0 & 0 & \Gamma_{32}^{S} + \Gamma_{31}^{S} & 0 & 0 & 0 \\
0 & 0 & 0 & 0 & 0 & 0 & \Gamma_{32}^{S} + \Gamma_{31}^{S} + \Gamma_{21}^{S} & 0 & 0 \\
0 & 0 & 0 & 0 & 0 & 0 & 0 & \Gamma_{32}^{S} + \Gamma_{31}^{S} + \Gamma_{21}^{S} & 0 \\
0 & 0 & 0 & - \dfrac{2}{\sqrt 3} \left( \Gamma_{21}^{S} + \Gamma_{32}^{S} - \Gamma_{31}^{S}  \right) & 0 & 0 & 0 & 0 &  2 \left( \Gamma_{32}^{S} + \Gamma_{31}^{S} \right) 
\end{pmatrix}
.
\end{equation}
\end{widetext}

It is worth to compare the obtained dissipative matrix $D^{(3)}_{kl}$ with those, that is usually used for experimental searches of the neutrino quantum decoherence $D_{\alpha \beta}=- \diag \left( \Upgamma_{21}, \Upgamma_{21}, \Upgamma_{3}, \Upgamma_{31}, \Upgamma_{31}, \Upgamma_{32}, \Upgamma_{32}, \Upgamma_{8} \right)$ where indices $\alpha, \beta = 1...8$ ($D_{0i}=D_{i0}=0$). Moreover, when analyzing the data on neutrino fluxes it is usually assumed that the decoherence parameters are equal to each other or to zero, (for example $\Upgamma_{12}=\Upgamma_{13}=\Upgamma_{23}$ or $\Upgamma_{12}=\Upgamma_{13}$, $\Upgamma_{23}=0$ \cite{DeRomeri:2023dht}).

The KamLAND experiment is sensitive only to decoherence parameter $\Upgamma_{21}$ (see, \cite{DeRomeri:2023dht,BalieiroGomes:2016ykp,Fogli:2007tx}, it means that other elements of dissipative matrix $D_{kl}$ does not play any crucial role. Therefore, it is possible to use experimental constraint on parameter $\Upgamma_{21}$ to limit the decay width $\Gamma_{21}^S$. 

Using the constraint from \cite{DeRomeri:2023dht} on $\Upgamma_{21}(E_0) < 1.8 \times 10^{-24}$ GeV (where $\Upgamma_{21}(E)=\Upgamma_{21}(E_0) \left(\dfrac{E}{E_0}\right)^{-1}$ with $E_0=1$ GeV) we obtain the constraint on the neutrino lifetime 

\begin{equation}\label{constrain}
    \dfrac{\tau_2}{m_2} = \dfrac{1}{2 \Upgamma_{21}(E_0) E_0} > 1.83 \times 10^{-10} \dfrac{s}{eV}
    .
\end{equation}
The existing constraints on visible neutrino decay from oscillating experiments are $\dfrac{\tau_2}{m_2}>6.7 \times 10^{-4} \dfrac{s}{eV}$ for Majorana case \cite{Funcke:2019grs} and $\alpha= E_\nu \Gamma^d <O(10^{-5} eV^2) $ \cite{Gago:2017zzy}. 

The direct constraint (\ref{constrain}) on the visible Dirac neutrino decay rate  from oscillation data obtained  for the first time.
Note that the obtained constraint (\ref{constrain}) requires a new more accurate data analysis with dissipative matrix (\ref{dis_matrix_3}).    

\section{Conclusion}

A new theoretical framework, based on the quantum field theory of open systems applied to neutrinos, has been developed. This framework aims to describe the neutrino evolution in an external environment, taking into account the effect of the neutrino quantum decoherence. Within this framework we have derived the new master equation (\ref{Main_Evoltion}) for the neutrino evolution that accounts for the neutrino decay into a lighter neutrino state and a massless scalar particle, as well as the inverse process of absorption of a massless particles by neutrino. Scalar particles (for example, axions), photons, gravitons and particles of higher spins are considered as a massless particle. 

The novelty of the derived generalize Lindblad master equation (\ref{Main_Evoltion}) in comparison to the commonly used Lindblad equation (\ref{Lindblad_eq}) is that it takes into account for the transitions between neutrino states with different neutrino momenta.

We also have considered a special case (\ref{special_equation}) of the derived master equation (\ref{Main_Evoltion}) when the neutrino decays on a massless axion for the case of neutrino degenerate hierarchy. This equation shows the applicability for the study of neutrino decays in fluxes from terrestrial sources and astrophysical sources. As an example, we have used the experimental results on neutrino decoherence parameter to constrain the neutrino lifetime $\dfrac{\tau_2}{m_2} > 1.83 \times 10^{-10} \dfrac{s}{eV}$.  

Due to the possibility for neutrino decays to be caused by physics beyond the Standard Model, the derived equation may provide a window into New physics.

\section*{Acknowledgement}
This work is supported in the framework of the State project ``Science'' by the Ministry of Science and Higher Education of the Russian Federation under the contract 075-15-2024-541.

\bibliography{main}

\end{document}